\documentclass[options]{JHEP3}
\usepackage{graphicx}
\newcommand{\vecp}[1]{\stackrel{\longrightarrow}{#1}}
\title{$\Lambda$ effect in the cosmological expansion of void}

\author{Henri-Hugues {\sc Fliche}\\
 LMMT\footnote{UPRES~EA 2596}, Fac. des Sciences et Techniques de St J\'er\^ome,
av. Normandie-Niemen, 13397 Marseille Cedex 20, France}
\author{Roland {\sc Triay}\\
 Centre de Physique Th\'eorique\footnote{Unit\'e Mixte de Recherche (UMR 6207) du CNRS, et des universit\'es Aix-Marseille I,
Aix-Marseille II et du Sud Toulon-Var. Laboratoire affili\'e \`a la FRUMAM (FR 2291).},
CNRS Luminy Case 907, 13288 Marseille Cedex 9, France\\ E-mail: 
\email{triay@cpt.univ-mrs.fr}}

\abstract{
We investigate the dynamical effect of the cosmological constant $\Lambda$ on a single spherical vacuum void evolving in the universe within a global solution of Newton-Friedmann models. As a result, the main characteristic is that the void expands with a huge initial burst up to match asymptotically the Hubble flow. The size of voids increases with $\Omega_{\circ}$ and with $\Lambda$, which is interpreted as respectively by the gravitational attraction of borders from outside regions and by the gravitational repulsion of vacuum from the inner region. The $\Lambda$-effect on the kinematics intervenes significantly by amplifying the expansion rate at redshift $z\sim 1.7$ for a background density parameter $\Omega_{\circ}\sim 0.3$. For a class of parameters values,  which corresponds in GR to spatially closed Friedmann models, it is interesting to note that a test particle in the inner region moves toward the border. Such a peculiar feature shows that the empty regions are swept out ; which stands as a stability criterion.
}
\preprint{JCAP/053P/0710}

\begin{document}
\section{Introduction}\label{Introduction}
 Foam like patterns with {\em large empty regions\/} are present in the distribution of observed galaxies at scales up to 100 Mpc (see {\it e.g.}, \cite{SoneiraPeebles78, JoeveerEinasto78, KirshnerEtal81,Einasto02, RojasEtal04, CrotonEtal04, JaanisteEtal04, ConroyEtal05}). Hence, one may wonder whether  the voids would be the most suitable structures to be studied for improving our understanding on the formation process of large cosmological structures. Such a question has motivated investigations which have been performed -- on their statistical properties, by improving identification techniques (see {\it e.g.}, \cite{SantiagoEtal06}), by exploring their formation process in a $\Lambda$CDM model through N-body simulations (see {\it e.g.}, \cite{GottloberEtal03, ShandarinEtal04, GoldbergVogeley04, Mathis04,Colberg04,WeygaertEtal04,PadillaEtal05}), and by probing their origins (see {\it e.g.}, \cite{OrdEtal05}); -- on the kinematics of giant voids (see {\it e.g.}, \cite{KopylovKopylova02}) ; -- and on their dynamics, by testing models of void formation (see {\it e.g.}, \cite{BenhamidoucheEtal99,FriedmannPiran00,Bolejko06}).

Herein, we propose to obtain a {\em global solution\/} describing the Newtonian dynamics of a spherical void in an expanding universe. The relevance of such a study in the context of cosmological structures can be justified as follows. From a theoretical point of view, it is clear that the dynamics of a single void does not allow to understand the evolution of a foam like structure but it is required as a first step of the investigate, however. From a standpoint of modeling observations, it appears that we can consider a homogeneous distribution of dust in which the void evolves. Indeed, at larger scales, according to observations, the dynamic of the cosmological expansion can be described by the Fried\-mann-Lema\^{\i}tre (FL) model, despite the presence of large scale structures. The reason can be found on theoretical grounds, due to the fact that the recession of galaxies together with the black body spectrum of Cosmological Background Radiation ensure that FL model provides us with a suitable description of the universe \cite{JMS74}.

Furthermore, the stability of FL models with respect to linear perturbations \cite{LifchitzKhalatnikov63} enable us to use perturbative approaches in order to understand the large scale structures formation process. Such investigations are generally performed with numerical simulations based on a FL cosmological background in which the structures are embedded. However, since voids are not linear structures \cite{BenhamidoucheEtal99}, a better understanding of the underlying dynamics requires a more adapted approach.  With this in mind and the aim of understanding the effect of a cosmological constant $\Lambda$ on the evolution of voids, we investigate the dynamics of {\em a single empty spherical shell\/} embeded in an uniformly distribution of dust which evolves with the cosmological expansion.

Our model is based on solutions of Euler-Poisson equations system for whom kinematics satisfy Hubble (cosmological) law. Except for the description of the gravitational waves, FL models can be described within such a framework (see {\it e.g.} \cite{McCrea55}) and Newtonian treatment of perturbations leads to similar results as those obtained in general relativity (GR) \cite{Fliche81}. Analogous problems have been investigated in GR \cite{Sato82, MaedaEtal82, MaedaSato83, SatoMaeda83, Maeda86} but with a different aim with respect to the cosmological constant effect on the dynamics of the void (empty bubble).

\section{Euler-Poisson equations system}
\label{Dynamics}
The  motion of a pressureless medium (dust)  is described by its specific density $\rho=\rho(\vec{r},t)$ and velocity $\vec{v}=\vec{v}(\vec{r},t)$ at position $\vec{r}$ and time $t$. These fields are constrained by Euler-Poisson-Newton (EPN) equations system: {\it i.e. \/} Euler equations system\footnote{Note that in the second Euler equation, we have
$\left(\frac{\partial\vec{v}}{\partial\vec{r}}\right)^{i}_{j}= \frac{\partial v^{i}}{\partial r^{j}}$.}
\begin{eqnarray}
\frac{\partial\rho}{\partial\,t}+{\rm div}\left(\rho \vec{v}\right)&=&0\label{Euler2}\\
\frac{{\rm d}\vec{v}}{{\rm d}t}=\frac{\partial \vec{v}}{\partial\,t}+\frac{\partial\vec{v}}{\partial\vec{r}}
\vec{v} &=& \vec{g} \label{Euler1}
\end{eqnarray}
where $\vec{g}=\vec{g}(\vec{r},t)$ stands for the gravitational field, and they satisfy the modified Poisson-Newton equations system
\begin{eqnarray}
\vecp{\rm rot}\vec{g} &=&\vec{0}\label{Poisson1}\\
{\rm div}\vec{g} &=&-4\pi{\rm G}\rho+\Lambda \label{Poisson2}
\end{eqnarray}
where ${\rm G}$ is Newton constant of gravitation. The sign of the cosmological constant  is chosen ($\Lambda\geq 0$) such that  the vacuum gravitational effect is repulsive. Accordingly to the status of gravitational field (which is not a vector field), we have chosen a cartesian coordinates system where $\vec{g}=\vec{0}$ at its origin, see \ref{Covariant}.

In order to disentangle the evolution of structures from the cosmological expansion, it is convenient to write these equations with new coordinates, herein named {\it reference coordinates\/}. They are defined as follows
\begin{equation}\label{tx}
(t,\quad \vec{x}=\frac{\vec{r}}{a}),\qquad a=a(t)>0,\qquad \dot{a}(t)\geq0,\qquad a_{\circ}=a(t_{\circ})=1
\end{equation}
where $a$ is a solution of Friedmann type differential equation
\begin{equation}\label{H}
\dot{H}+H^{2}-\frac{\Lambda}{3}+\frac{4\pi{\rm G}}{3}\rho_{\circ}a^{-3}=0,\qquad H=\frac{\dot{a}}{a}
\end{equation}
and $\rho_{\circ}$ stands for an arbitrary constant; herein, the dotted variables stand for time derivatives. This differential equation admits a integration constant
\begin{equation}\label{K}
K_{\circ}=\frac{8\pi{\rm G}}{3} \rho_{\circ}+\frac{\Lambda}{3}-H_{\circ}^{2},\qquad
H_{\circ}=H(a_{\circ})
\end{equation}
The integration of Eq.\,(\ref{H}) provides us with Friedmann equation
\begin{equation}\label{chrono}
H^{2}=\frac{\Lambda}{3}-\frac{K_{\circ}}{a^{2}}+\frac{8\pi{\rm G}}{3} \frac{\rho_{\circ}}{a^{3}}\geq 0,\qquad
H_{\infty}=\lim_{a\to\infty}H=\sqrt{\frac{\Lambda}{3}}
\end{equation}
where $H_{\infty}$ is the asymptotical value of the function $H(a)$ when $\Lambda>0$.

The differential equation given in Eq.\,(\ref{chrono}) is integrated by a quadrature, and the function $a(t)$ is obtained as a reciprocal mapping. It depends on three parameters chosen among $\Lambda$, $\rho_{\circ}$, $H_{\circ}$ and $K_{\circ}$, see Eq.\,(\ref{K}). At this stage, it is important to mention that, excepted $\Lambda$, they do not identify (yet) with the cosmological parameters of FL model but they intervene solely in the choice of coordinates system. EPN equations system Eq.\,(\ref{Euler2}--\ref{Poisson2}),  reads in term of reference coordinates as follows
\begin{eqnarray}
\frac{\partial\rho_{\rm c}}{\partial\,t}+{\rm div}\left(\rho_{\rm c} \vec{v}_{\rm c}\right)=0\label{Euler2n}\\
\frac{\partial \vec{v}_{\rm c}}{\partial\,t}+\frac{\partial\vec{v}_{\rm c}}{\partial\vec{x}} \vec{v}_{\rm c} +2H\vec{v}_{\rm c} = \vec{g}_{\rm c} \label{Euler1n}\\
\vecp{\rm rot}\vec{g}_{\rm c} =\vec{0}\label{Poisson1n}\\
{\rm div}\vec{g}_{\rm c}  =-\frac{4\pi{\rm G}}{a^{3}}\left(\rho_{\rm c} -\rho_{\circ}\right)\label{Poisson2n}
\end{eqnarray}
where ``${\rm div}$'' and ``$\vecp{\rm rot}$'' denote the differential operators with respect to variable $\vec{x}$,
\begin{eqnarray}\label{VetG}
\rho_{\rm c}=\rho a^{3},\qquad
\vec{v}_{\rm c}=\frac{{\rm d}\vec{x}}{{\rm d}t}
\end{eqnarray}
act as the density and the velocity fields of the medium in the {\em reference frame\/}, and
\begin{equation}\label{gC}
\vec{g}_{\rm c}=\frac{\vec{g}}{a}+\left(\frac{4\pi{\rm G}}{3a^{3}}\rho_{\circ}- \frac{\Lambda}{3}\right)\vec{x}
\end{equation}
as the acceleration field.

\subsection{Newton-Friedmann and Vacuum models}\label{solutions}
In the reference frame, a uniform distribution of dust and an empty space are described, as obvious solutions of Eq.~(\ref{Euler2n},\ref{Euler1n},\ref{Poisson1n},\ref{Poisson2n}), respectively by~:
\begin{enumerate}
\item  a Newton-Friedmann model (NF), which is defined by 
\begin{equation}\label{FL}
\rho_{\rm c}=\rho_{\circ},\qquad
\vec{v}_{\rm c}=\vec{0}, \qquad
\vec{g}_{\rm c}=\vec{0}
\end{equation}
Herein, according to Eq.\,(\ref{chrono},\ref{VetG}), $a$ and $H$ recover their usual interpretations as {\em expansion parameter\/} and {\em Hubble parameter\/} respectively, $\rho_{\circ}=\rho a^{3}$ identifies to the density of sources in a comoving space (of scalar curvature $K_{\circ}$). We focus our investigation to cosmological expansions  which excludes bouncing solutions, {\it i.e.\/}, $\dot{a}\geq 0$. Hence, an analysis on roots of third degree polynomials shows that the condition
\begin{equation}\label{contraint}
K_{\circ}^{3}<\left(4\pi{\rm G}\rho_{\circ}\right)^{2}\Lambda
\end{equation}
must be fulfilled. From primordial epochs, $H$ decreases with time to reach its asymptotic value $H_{\infty}$. However, the kinematics has two distinct behaviors, characterized by the sign of $K_{\circ}$: if $K_{\circ}\leq 0$ then $H$ decreases monotonically, but if $K_{\circ}>0$ then it reaches first a minimum $H_{m}$ from where it grows to reach $H_{\infty}$ from downward. This minimum is defined by
 \begin{equation}\label{extremum}
H_{m}=H_{\infty}\sqrt{1-\frac{K_{\circ}^{3}}{\Lambda\left(4\pi{\rm G} \rho_{\circ}\right)^{2}}}<H_{\infty}
\end{equation}
at $a=4\pi{\rm G}\rho_{\circ}K_{\circ}^{-1}$, it characterizes a  {\it loitering period\/}. Let us remind that FL models are characterized by the following dimensionless parameters
\begin{equation}\label{CP}
\lambda_{\circ} =\frac{\Lambda}{3H_{\circ} ^{2}},\qquad
\Omega_{\circ} = \frac{8\pi {\rm G}\rho_{\circ}}{3H_{\circ}^{2}},\qquad
k_{\circ} = \frac{K_{\circ}}{H_{\circ} ^{2}}=\lambda_{\circ}+\Omega_{\circ}-1
\end{equation}
These notations are preferred to the usual $\Omega_{\Lambda}=\lambda_{\circ}$ and $\Omega_{K}=-k_{\circ}$ for avoiding ambiguities on the interpretation of cosmological parameters, see {\it e.g.\/}, \cite{FlicheSouriau79},\cite{FlicheEtal82}. A dimensional analysis of Eq.\,(\ref{K}) shows that the Newtonian interpretation of $k_{\circ}$ corresponds to a {\em dimensionless binding energy\/} for the universe during its expansion; the lower $k_{\circ}$ the faster the cosmological expansion.  In the FL world model,  $k_{\circ}$ stands for the dimensionless spatial curvature of the comoving space; $k_{\circ}>0$ characterizes a spatially closed universe.
 
\item  a Vacuum model (V), which is defined by
\begin{equation}\label{void}
\rho_{\rm c}=0,\qquad
\vec{v}_{\rm c}=\left( H_{\infty}-H \right) \vec{x}, \qquad
\vec{g}_{\rm c}=\frac{4\pi{\rm G}}{a^{3}}\rho_{\circ}\vec{x}
\end{equation}
\end{enumerate}

\section{Spherical  voids in Newton-Friedman universe}\label{DynamicsVoid}
For avoiding misunderstandings in the interpretation of the dynamics, we use a covariant formulation of Euler-Poisson equations system (see~\ref{Covariant}). The model of a spherical void embedded in a uniform distribution of dust is obtained by sticking together the V and NF models, both as local solutions of EPN equations system defined in Sec.~\ref{solutions}. It accounts for the dynamics of their common border ({\it i.e.\/} the boundaries conditions), which is assumed to be a {\it shell\/} of null thickness.

\subsection{Dynamical model}\label{ModelVoid}
We consider three distinct media~: a spherical thin shell (S) on which dust is uniformly distributed, an empty space (V) inside, and a uniform distribution of dust (NF) outside. For convenience in writing, the symbols S, V and NF denote respectively the related medium and the corresponding model. They behave such that S makes the juncture of V with NF according to Eq.\,(\ref{FL},\ref{void}). The tension-stress on S is assumed to be negligible, which is characterised  by a (symmetric contravariant) stress-energy tensor defined as follows
\begin{equation}\label{Ts}
T_{\rm S}^{00}=(\rho_{\rm S})_{c},\quad
T_{\rm S}^{0j}=(\rho_{\rm S})_{c}v_{c}^{j},\quad
T_{\rm S}^{jk}=(\rho_{\rm S})_{c}v_{c}^{j}v_{c}^{k}
\end{equation}
The background is described by the following stress-energy tensor
\begin{equation}\label{To}
T_{\rm NF}^{00}=\rho_{c},\quad
T_{\rm NF}^{0j}=0,\quad
T_{\rm NF}^{jk}=0
\end{equation}
Because the eulerian functional
\begin{equation}\label{T}
{\cal T}(x\mapsto\gamma)=\int T_{\rm S}^{\mu\nu} \gamma_{\mu\nu}{\rm d}t{\rm d}S+
\int T_{\rm NF}^{\mu\nu} \gamma_{\mu\nu}{\rm d}t{\rm d}V
\end{equation}
vanishes when $\gamma$ reads in the form $\gamma_{\mu\nu}=\frac{1}{2}\left( \hat{\partial}_{\mu} \xi_{\nu}+ \hat{\partial}_{\nu} \xi_{\mu}\right)$, one has
\begin{eqnarray}
&&\int_{\rm S}
\left(\left(\partial_{0}\xi_{0}+g_{c}^{j}\xi_{j}\right)+\left(\partial_{j}\xi_{0}+\partial_{0}\xi_{j}-2H\xi_{j}\right)v_{c}^{j} 
+ v_{c}^{j}v_{c}^{k}\partial_{j}\xi_{k}\right)(\rho_{\rm S})_{c}\,{\rm d}t x^{2}{\rm d}\Omega \nonumber\\
&&=- \int_{\rm NF} \rho_{c}\partial_{0}\xi_{0}\,{\rm d}t  x^{2} {\rm d}x{\rm d}\Omega \label{EqFond}
\end{eqnarray}
where ${\rm d}\Omega$ stands for the solid angle element (see~\ref{Covariant}). The radial symmetry of solutions enables us to write the {\em reduced\/} peculiar velocity and acceleration of a test particle located on the shell as follows
\begin{equation}\label{radial}
\vec{v}_{c}=\alpha\vec{x},\qquad
\vec{g}_{c}=\beta\vec{x}
\end{equation}
where the functions $\alpha=\alpha(t)$ and $\beta=\beta(t)$ must be determined. A by part integration of Eq.\,( \ref{EqFond}) provides us with
\begin{eqnarray}\label{EulerS}
&&
\int_{t_{1}}^{t_{2}}
\left(\partial_{0}(\rho_{\rm S})_{c}+3(\rho_{\rm S})_{c}\alpha-\rho_{c} \alpha x\right)x^{2}\xi_{0} {\rm d}t\\
&=&
\int_{t_{1}}^{t_{2}}
\left(\partial_{0}\left((\rho_{\rm S})_{c}\alpha\right)  +4(\rho_{\rm S})_{c}\alpha^{2} + 2H(\rho_{\rm S})_{c}\alpha-
(\rho_{\rm S})_{c}\beta x^{-1}\right)x^{3}\tilde{\xi}
{\rm d}t\nonumber
\end{eqnarray}
where $x=\|\vec{x}\|$ stands for the radius of S and $\tilde{\xi}=\sqrt{\xi_{1}^{2}+\xi_{2}^{2}+\xi_{2}^{2}}$.  This equality must be fulfilled for all bounded time intervals and compact support 1-forms. Hence, we easily derive the conservation equations for the mass
\begin{equation}\label{masseC}
\partial_{0}(\rho_{\rm S})_{c}+\left(3(\rho_{\rm S})_{c}-\rho_{c} x\right) \alpha = 0
\end{equation}
and for the momentum
\begin{equation}\label{impulsionC}
\frac{{\rm d}\alpha}{{\rm d}t}+\left(1+\frac{\rho_{c}}{(\rho_{\rm S})_{c}}x\right)\alpha^{2}+2H \alpha+ \frac{\beta}{x}=0
\end{equation}
With Eq.\,(\ref{tx}), the calculation of the gravitational force from the entire shell acting on a particular point\footnote{The modified newtonian gravitation field reads
$
\vec{g}=\left(
\frac{\Lambda}{3}-\frac{{\rm G}m}{r^{3}}\right)\vec{r}
$}
provides us with
\begin{equation}\label{beta}
\beta=\frac{4\pi{\rm G}}{a^{3}}\left(\frac{\rho_{c}}{3}- \frac{(\rho_{\rm S})_{c}}{2x} \right)
\end{equation}
It is noticeable that $(\rho_{\rm S})_{c}=\rho_{c}x/3$ ensures that the mass forming the shell comes from its interior\footnote{This property agrees with spherical voids originating from a uniformly distributed matter, although other behaviors can be envisaged. Indeed, Eq.\,(\ref{masseC}) reads
$$
\partial_{0}\delta+3 \alpha \delta = 0$$
where
$$
\delta= (\rho_{\rm S})_{c}-\frac{1}{3}\rho_{c} x=f(x_{i})\exp{\left(-3\int_{t_{i}}^{t}\alpha{\rm d}t\right)}
$$
accounts for the excess/defauft of the shell density with respect to that formed from an uniform distribution of matter, $f(x_{i})$ stands for its initial value at $t=t_{i}$.}. It must be understood as a particular solution of Eq.\,(\ref{masseC}) which is compatible with initial conditions provided by FL model. By focusing our investigation to such a model, Eq.\,(\ref{impulsionC}) transforms
\begin{equation}\label{impulsionC1}
\frac{{\rm d}\alpha}{{\rm d}t}+4\alpha^{2}+2H \alpha- \frac{2\pi{\rm G}}{3} \frac{\rho_{c}}{a^{3}}=0
\end{equation}
With the dimensionless variable
\begin{equation}\label{chi}
\chi=4\frac{\alpha}{H_{\circ}}a^{2}
\end{equation}
where the ratio $\alpha\,H_{\circ}^{-1}$ stands for the expansion rate of S in the reference frame, Eq.\,(\ref{impulsionC1}) tranforms into a Riccati equation
\begin{equation}\label{impulsionC2}
\frac{{\rm d} \chi}{{\rm d}a}= \left(\Omega_{\circ}-\frac{\chi^{2}}{a} \right)
\frac{1}{\sqrt{P(a)}}
\end{equation}
where
\begin{equation}\label{P}
P(a)= \lambda_{\circ}a^{4}-k_{\circ}a^{2}+\Omega_{\circ} a,\qquad P(1)=1
\end{equation}
According to Eq.\,(\ref{VetG},\ref{chi}), one has
\begin{equation}\label{Croissance}
x=x_{i}\exp\left(\int_{a_{i}}^{a}\frac{\chi{\rm d}a}{4a\sqrt{P(a)}}
\right)
\end{equation}
where $x_{i}=x(t_{i})$ and $a_{i}=a(t_{i})$ stands for initial values (at time $t=t_{i}$).

\subsection{Qualitative analysis}\label{Analysis}
A meaningful qualitative analysis of the expansion of the void must be performed by means of dimensionless quantities. Herein, we chose the {\it magnification\/} $X$ and the {\it expansion rate\/} $Y$, which are defined as follows
\begin{equation}\label{XY}
X=\frac{x}{x_{i}},\qquad Y=\frac{\alpha}{H_{\circ}}
\end{equation}
Their diagrams versus the expansion parameter $a$ characterize adequately the expansion dynamics of S; they are obtained from Eq.\,(\ref{impulsionC2}, \ref{Croissance}) by numerical integration. They can be easily translated in term of cosmic time $t$ since $a(t)$ is a monotonic function of $t$  in the present investigation. The initial conditions lie on the expansion rate $Y_{i}$ and the formation date $t_{i}$, as expressed by means of $a_{i}=a(t_{i})$. Herein, we investigate voids that initially expand with Hubble flow $Y_{\rm i}=0$ at $a_{\rm i}=0.003$. Instead of having an exhaustive analysis with respect to cosmological parameters, we limit our investigation around the generally accepted values $\lambda_{\circ}=0.7$ and $\Omega_{\circ}=0.3$.

The kinematics of the expansion is discussed in Sec.\,\ref{kinematics}, the dependence on cosmological parameters (herein named $\Omega$-effect and $\Lambda$-effect respectively) in Sec.\, \ref{dependence}, the role of initial conditions in Sec.\, \ref{InitCind},  and the results are synthesised in Sec.\,\ref{Synthesis}. 

\subsubsection{Kinematics}\label{kinematics}
\begin{figure}[htbp]
\begin{center}
\includegraphics[width=3.5in]{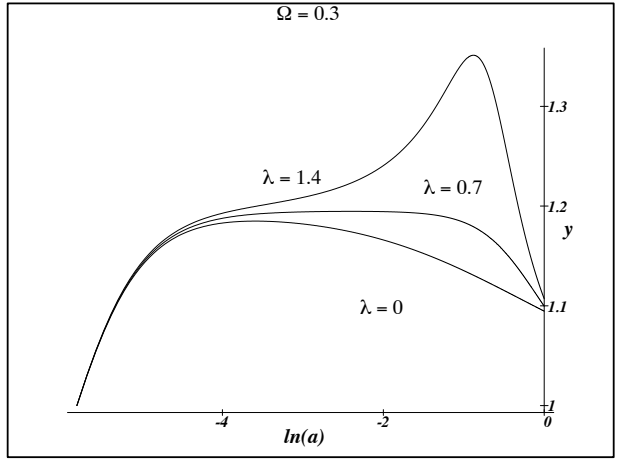}
\caption{The corrective factor to Hubble expansion.}
\label{Fig0}
\end{center}
\end{figure}

The expansion velocity of the shell S with respect to its centre reads 
\begin{equation}\label{vitesse}
\vec{v}=y H \vec{r},\qquad y=1+\frac{Y}{h},\qquad h=\frac{H}{H_{\circ}}
\end{equation}
where $y$ stands for the corrective factor  to Hubble expansion.  According to Fig.\,\ref{Fig0}, which shows $y$ versus $a$ for  $\Omega_{\circ}=0.3$ and for three values of $\lambda_{\circ}\in\{0, 0.7, 1.4\}$, S expands faster than Hubble expansion ({\it i.e.\/}, $y>1$) all along its evolution. It starts with a huge increase of the velocity at early stages (such as a burst) with no significant dependence on $\Lambda$, and then it decreases for reaching asymptotically the Hubble flow\footnote{Although the present epoch ($a=1$) appears quite peculiar because of the relative proximity of these curves, this turns out to be solely an artefact since they (will) cross at $a>1$ but not in a single point.}.
The $\Lambda$ effect is to delay the decreasing of the expansion velocity of S.  On the diagram, it is characterised by a protuberance on the curves at $z\sim 1.7$ for $\lambda_{\circ}>0.7$ ({\it i.e.\/}, $k_{\circ}>0$) , the larger the $\lambda_{\circ}$ ({\it i.e.\/}, $k_{\circ}$) the higher the bump. Such a feature is due to the existence of a minimum value  $H_{m}$ of Hubble parameter $H$, which is reached during the cosmological expansion and is related to a coasting period, see Eq.\,(\ref{extremum}). With respect to observation of distant voids, the Doppler shift between the edge and the center of a spherical void which is observed at redshift $z$ does not exceed 
\begin{equation}\label{redshift}
\Delta\,z= \frac{1}{(1+z)} \frac{v_{c}}{c}=\frac{XY}{1+z} \frac{x_{i}H_{\circ}}{c}
\end{equation}
where $c$ is the speed of the light.

\subsubsection{Dichotomy between cosmological parameters}\label{dependence}
The growth of spherical voids is investigated by means of X and Y diagrams with respect to cosmological parameters $\Omega_{\circ}$ and  $\lambda_{\circ}$, with the aim of disentangling their related effects.
\begin{enumerate}
\item $\Omega$-effect.\label{OmegaVoid} ---
The dependence on the outward density is analyzed with a vanishing cosmological constant $\lambda_{\circ}=0$ and three values for the density parameter $\Omega_{\circ}\in\{0.15,0.3,0.45\}$. The comparison of diagrams in Fig.\,\ref{Fig1} tell us that  the larger the $\Omega_{\circ}$ the larger the magnification $X$.
\begin{figure}[htbp]
\begin{center}
\includegraphics[width=3.5in]{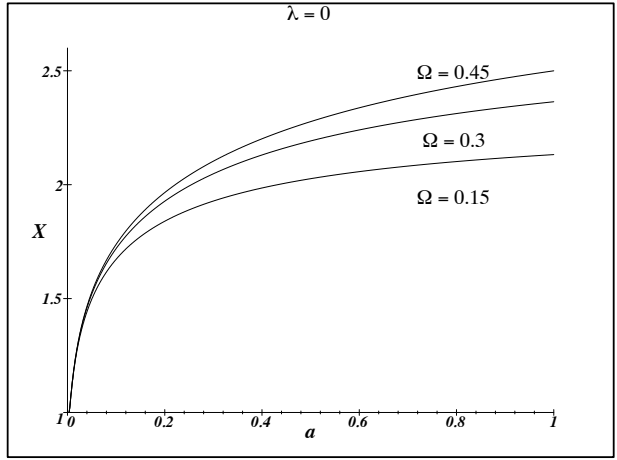}
\caption{The magnification $X$ --- dependence on density.}
\label{Fig1}
\end{center}
\end{figure}
Such an effect results from the attraction of shell particles toward denser regions. 
\begin{figure}[htbp]
\begin{center}
\includegraphics[width=3.5in]{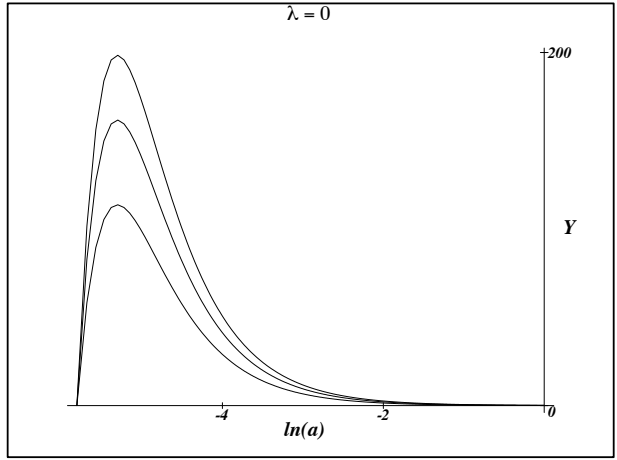}
\caption{The expansion rate $Y$ --- dependence on density.}
\label{Fig2}
\end{center}
\end{figure}
According to Fig.\,\ref{Fig2}, the growth looks like a huge burst which freezes asymptotically up to matching Hubble expansion ($Y=0$). Moreover, the larger the $\Omega_{\circ}$ the larger the expansion rate. This trend is not significantly modified by other acceptable values of $\lambda_{\circ}$.

 \item $\Lambda$-effect.\label{LambdaVoid} ---
The dependence on the cosmological constant is analyzed at constant $\Omega_{\circ}=0.3$ and $\lambda_{\circ}\in\{0,0.7,1.4\}$. The comparison of diagrams 
in Fig.\,\ref{Fig3} shows that  the larger the $\lambda_{\circ}$ the larger the magnification $X$, which increases nonlinearly with $\lambda_{\circ}$.
\begin{figure}[htbp]
\begin{center}
\includegraphics[width=3.5in]{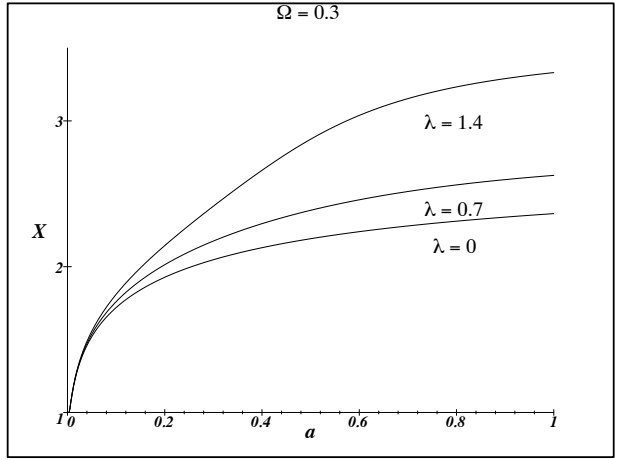}
\caption{The magnification $X$ --- dependence on the cosmological constant.}
\label{Fig3}
\end{center}
\end{figure}
\begin{figure}[htbp]
\begin{center}
\includegraphics[width=3.5in]{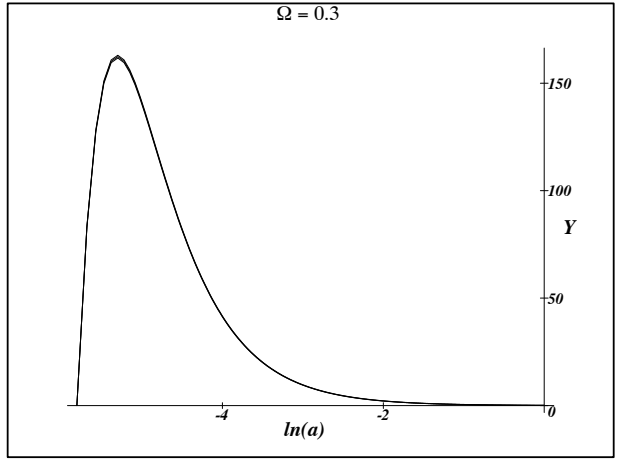}
\caption{The expansion rate $Y$ --- dependence on the cosmological constant.}
\label{Fig4}
\end{center}
\end{figure}
According to Fig.\,\ref{Fig4},  the expansion rate $Y$ does not characterize $\Lambda$ since the related curves do not disentangle. Its dependence on $\Lambda$ is weak and appears later on the evolution, but it has a cumulative effect which is reflected on $X$, see Fig.~\ref{Fig0}.
\end{enumerate}
According to Eq.\,(\ref{K}), the dependence of the dynamics of S on the parameter $k_{\circ}$ can be retrieved from above analysis; the higher $k_{\circ}$ the faster the S expansion, and a similar effect for the magnification. Let us point out an interesting feature that characterizes the sign of $k_{\circ}$ which appears on the motion  in the reference frame of a test particle inside S. According to Eq.\,(\ref{void})~:
\begin{itemize}
\item if $k_{\circ}\leq 0$ then the test particle moves toward the centre of the void;
\item if $k_{\circ}>$ then it starts moving toward its border at date $t=t^{\star}$ defined such that $a(t^{\star})=\frac{8}{3} \pi{\rm G}\rho_{\circ}K_{\circ}^{-1}$, which happens when Hubble expansion $H<H_{\infty}$. This feature, of sweeping out the inner part of S toward its border, stands for a stability criterion for voids which acts only for a spatially closed universe. 
\end{itemize}

\subsubsection{Dependence on initial conditions}\label{InitCind} 
\begin{figure}[htbp]
\begin{center}
\includegraphics[width=3.5in]{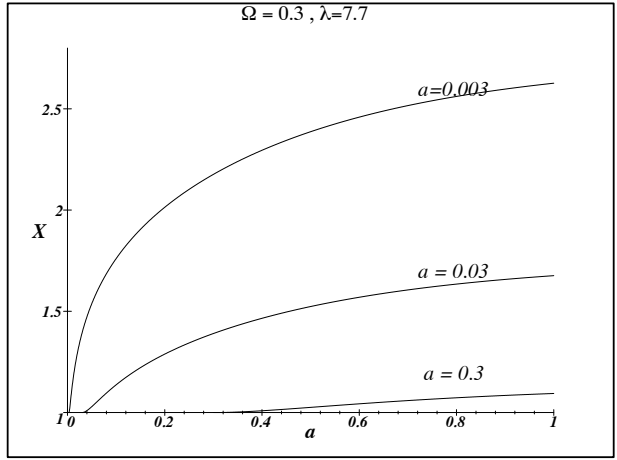}
\caption{The magnification $X$ --- dependence on the formation date.}
\label{Fig5}
\end{center}
\end{figure}

---
Figure\,\ref{Fig5} shows three curves corresponding to formation dates $a_{i}\in\{0.003, 0.03,0.3\}$ by assuming $\Omega_{\circ}=0.3$, $\lambda_{\circ}=0.7$ and with  $Y_{i}=0$.  As expected, the earlier the date of birth the larger the magnification $X$.

\begin{figure}[htbp]
\begin{center}
\includegraphics[width=3.5in]{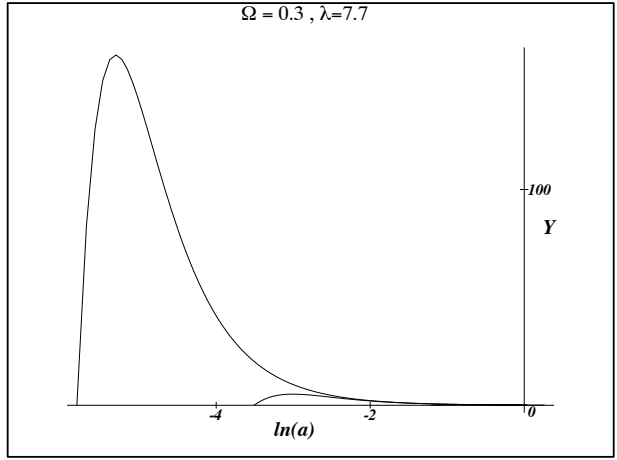}
\caption{The expansion rate $Y$ --- dependence on the formation date.}
\label{Fig6}
\end{center}
\end{figure}

Figure\,\ref{Fig6}  suggests the existence of a limiting curve $a\mapsto Y$ defined by $a_{i}=0$, which characterises $\Omega_{\circ}$. The expansion rate curves related to other formation dates are located in lower part and reach it asymptotically. 
One can expect that the effects on the dynamics of growth resulting from reasonable initial expansion rates $Y_{i}\neq 0$, related to physical processes ({\it e.g.\/}, supernovae explosions \cite {OstrikerCowie81}), are negligible at primordial epoch\footnote{Since Hubble expansion is all the more important towards the past, the earlier the formation date the weaker this effect, according to Eq.\,(\ref{vitesse}).}, what legitimises the initial condition $Y_{i}=0$ at $a_{i}=0.003$. Evolutions with other initial conditions on $Y_{i}$ can be deduced from that since any given point $(a_{i},Y_{i})$ in the diagram belongs to a single evolution curve.

\subsubsection{Synthesis}\label{Synthesis}
-- As a result, the expansion of a spherical empty region does not show a  {\it linear regime\/}, but it starts with a huge burst which freezes asymptotically up to match Hubble expansion. Its radius  (in the reference frame) $x$ increases from an initial size $x_{i}$ with the cosmological parameters $\Omega_{\circ}$ and $\lambda_{\circ}$. The related individual effects can be interpreted respectively by the gravitational attraction from the outer parts and by the gravitational repulsion of its borders (or equivalently, the vacuum). The larger these parameters  the higher the magnification $X=x/x_{i}$. The dynamics is sensitive to $\Omega$-effect at primordial epochs and to $\Lambda$-effect later on. The evolution of its expansion velocity  with time in the reference frame $\vec{v}_{c}=YH_{\circ}\vec{x}$ does not characterise $\lambda_{\circ}$ but $\Omega_{\circ}$. The cosmological constant intervenes mainly by delaying the decrease of the velocity expansion. The expansion rate evolutions related to formation dates show a common envelope curve which characterises the density parameter $\Omega_{\circ}$.
 
\subsection{Discussion}\label{Discussion}
It must be kept in mind that, based on two exact solutions of Euler-Poisson equations system, this model allows one to investigate properly the behaviour of a single spherical void. The connection conditions between the inner and the outer regions, which are investigated in a (covariant) Classical Mechanics, provides us unambiguously with the dynamics of the shell. Such a schema does not correspond to an embedding of a void in a cosmological background solution (as usual), but it stands for a non linear approach of NF models. However, while it can be useful in N-bodies simulations to improve voids dynamics, it requires stringent properties on the space distribution of neighboring matter and ignores GR effects. At this step, one expects that it can apply solely to spherical voids with reasonable size in order to guarantee reliable results.

With the aim of using the characteristics of voids to determine the values of the cosmological parameters, it is interesting to mention that the property of $K_{\circ}$ of being an integration constant (see Eq.~\ref{K}, \ref{CP}) enables us to use $k_{\circ}$ to classify the dynamics of the cosmological expansion. With this in mind, let us note that the spatially closed world models $k_{\circ}>0$ are characterized by a significant expansion of voids which reaches its maximum at redshift $z\sim 1.7$ (with $\Omega_{\circ}\sim 0.3$), the larger the $\lambda_{\circ}$ ({\it i.e.\/}, $k_{\circ}$) the higher the expansion rate. The inner part of voids shows a de Sitter expansion $\vec{v}=\sqrt{\Lambda/3}\vec{r}$ which sweeps out it, what can be interpreted as a stability criterion. 

\section{Conclusion}\label{Conclusion}
Using Euler-Poisson equations system with cosmological constant, we investigate the dynamics of an isotropic universe constituted by a spherical (empty) void surrounded by a uniform distribution of dust. It turns out that the behaviours of both regions (inside and outside the shell) coincide with Newton-Friedmann (NF) solutions, which ensures a consistent model. The dynamics of the void expansion is analyzed in the reference (comoving) frame

The general behaviour of the void expansion shows a huge initial burst, which freezes asymptotically up to match Hubble expansion. While the corrective factor to Hubble law on the shell depends weakly on the cosmological constant $\Lambda$ at early stages, it enables us to disentangle significantly cosmological models around redshift $z\sim 1.7$. The magnification of spherical voids increases with the cosmological parameters related to the density and to the cosmological constant. In case of spatially closed Friedmann models, the empty regions are swept out, which provides us with a stability criterion. It could be applied to observational data, as long as the matter distribution in the outer region of the single void is reasonably uniform

\appendix
\section{Covariant formulation of Euler-Poisson equations system}\label{Covariant}
Souriau's covariant formulation of Euler-Poisson equations\cite{Souriau70} can be summarized as follows. The geometrical interpretation of Newton dynamics\cite{DuvalKunzle78} shows that the component of gravitational field $\vec{g}$ can be identified to the unique non vanishing Christoffel components of newtonian connexion 
$$
\Gamma^{j}_{00}=-g^{j}
$$
Hence, one obtains their expression in the new coordinates system defined in Eq.\,(\ref{tx}),  and the only non null component reads
$$
(\Gamma_{c})^{j}_{k0}=H \delta^{j}_{k},\qquad (\Gamma_{c})^{j}_{00}=-g_{c}^{j}
$$
Let ${\cal T}$ be a functional\footnote{A functional is a real-valued function on a vector space, usually of functions. It is often used in the variational calculus, in particular in Mechanics.} defined on the set of symmetric density tensors $x\mapsto \gamma$ on newtonian spacetime $\mathbf{R}^{4}$
$$
{\cal T}(x\mapsto\gamma)=\int_{\mathbf{R}^{4}}T^{\mu\nu} \gamma_{\mu\nu}{\rm d}t{\rm d}V
$$
where ${\rm d}t{\rm d}V$ stands for the volume element and $T$ for a symmetric contravariant tensors  which accounts for the media. The measure density ${\cal T}$ is eulerian if and only if it vanishes for all covariant tensor fields that reads
$$
\gamma_{\mu\nu}=\frac{1}{2}\left( \frac{\hat{\partial} \xi_{\nu}}{\partial x^{\mu}}+ \frac{\hat{\partial} \xi_{\mu}}{\partial x^{\nu}}\right)
$$
where $\hat{\partial}$ stands for the covariant derivative\footnote{Let us remind the definition of the covariant derivative $\frac{\hat{\partial} \xi_{\nu}}{\partial x^{\mu}}=\frac{\partial \xi_{\nu}}{\partial x^{\mu}}+ \Gamma^{\lambda}_{\mu\nu}\xi_{\lambda}$} and $x\mapsto\xi$ for a \underline{compact support} 1-form.
In such a case, it is obvious to show that 
$$
\frac{\hat{\partial}T^{\mu\nu}}{\partial x^{\mu}}=0
$$
can be interpreted as the Euler equations, in any coordinates systems.

%\section*{References}

\end{document}